\documentclass[12pt,english]{article}
\pdfoutput=1
\usepackage{lmodern}

\usepackage[T1]{fontenc}
\usepackage[latin9]{inputenc}
\usepackage{geometry}
\usepackage{color}
\usepackage{babel}
\usepackage{cite}
\usepackage{mathtools}
\usepackage{amsmath}
\usepackage{amssymb}
\usepackage{graphicx}
\usepackage{esint}
\usepackage{bbm}
\usepackage[unicode=true,pdfusetitle,
 bookmarks=true,bookmarksnumbered=false,bookmarksopen=false,
 breaklinks=false,pdfborder={0 0 1},backref=false,colorlinks=true]
 {hyperref}
\hypersetup{
 citecolor=black,linkcolor=black,urlcolor=black}
\makeatletter

 \@ifundefined{textcolor}{}
 {
   \definecolor{BLACK}{gray}{0}
   \definecolor{WHITE}{gray}{1}
   \definecolor{RED}{rgb}{1,0,0}
   \definecolor{GREEN}{rgb}{0,1,0}
   \definecolor{BLUE}{rgb}{0,0,1}
   \definecolor{CYAN}{cmyk}{1,0,0,0}
   \definecolor{MAGENTA}{cmyk}{0,1,0,0}
   \definecolor{YELLOW}{cmyk}{0,0,1,0}
 }
\usepackage{babel}
\makeatletter
\def\simgt{\mathrel{\lower2.5pt\vbox{\lineskip=0pt\baselineskip=0pt
           \hbox{$>$}\hbox{$\sim$}}}}
\def\simlt{\mathrel{\lower2.5pt\vbox{\lineskip=0pt\baselineskip=0pt
           \hbox{$<$}\hbox{$\sim$}}}}
\makeatother

\newcommand{\be}{\begin{equation}}
\newcommand{\ee}{\end{equation}}
\newcommand{\bea}{\begin{eqnarray}}
\newcommand{\eea}{\end{eqnarray}}

\newcommand{\Ref}[1]{Ref.~\cite{#1}}
\newcommand{\Fig}[1]{Fig.~\ref{#1}}
\newcommand{\Eq}[1]{Eq.~\eqref{#1}}
\newcommand{\Eqs}[2]{Eqs.~\eqref{#1} and \eqref{#2}}
\newcommand{\Sec}[1]{Sec.~\ref{#1}}

\newcommand{\stacklines}[2]{\genfrac{}{}{0pt}{}{#1}{#2}}

\begin{document}

\interfootnotelinepenalty=10000
\baselineskip=18pt

\hfill CALT-TH-2017-20
\hfill

\vspace{2cm}
\thispagestyle{empty}
\begin{center}
{\LARGE \bf
Hidden Simplicity of the Gravity Action
}\\
\bigskip\vspace{1cm}{
{\large Clifford Cheung and Grant N. Remmen}
} \\[7mm]
 {\it Walter Burke Institute for Theoretical Physics \\[-1mm]
    California Institute of Technology, Pasadena, CA 91125} \let\thefootnote\relax\footnote{e-mail:
\url{clifford.cheung@caltech.edu,gremmen@theory.caltech.edu}} \\
 \end{center}
\bigskip
\centerline{\large\bf Abstract}

\begin{quote} \small

We derive new representations of the Einstein-Hilbert action in which graviton perturbation theory is immensely simplified.  To accomplish this, we recast the Einstein-Hilbert action as a theory of purely cubic interactions among gravitons and a single auxiliary field.  The corresponding equations of motion are the Einstein field equations rewritten as two coupled first-order differential equations.  Since all Feynman diagrams are cubic, we are able to derive new off-shell recursion relations for tree-level graviton scattering amplitudes. With a judicious choice of gauge fixing, we then construct an especially compact form for the Einstein-Hilbert action in which all graviton interactions are simply proportional to the graviton kinetic term.  Our results apply to graviton perturbations about an arbitrary curved background spacetime.

\end{quote}
\newpage

\tableofcontents

\newpage

\setcounter{footnote}{0}

\section{Introduction}

The modern scattering amplitudes program has revealed a striking simplicity in gravity that suggests an underlying structure not yet fully understood. At the same time, groundbreaking progress on the experimental front of gravitational wave astronomy \cite{LIGO,Abbott:2016nmj,Abbott:2017vtc} has created new opportunities for utilizing insights from formal theory.  It is therefore critical that we fully appraise to what extent theoretical advances in gravity have anything to offer by way of real-world applications.

In this paper, we present alternative representations of the Einstein-Hilbert (EH) action that exhibit an immense reduction in the complexity of graviton perturbation theory.    Our results involve a general procedure for integrating in auxiliary fields to recast the EH action into a form in which all interactions truncate at finite order.  In the minimal construction presented in this paper, we expand the EH action about a flat background in terms of graviton perturbations $h_{ab}$ and a single auxiliary field $A^a_{bc}=A^a_{cb}$ interacting via purely cubic vertices,
\be
S_{\rm EH} = -\frac{1}{16\pi G} \int \mathrm{d}^D x \left[\left(A^a_{bc} A^b_{ad} - \frac{1}{D-1} A^a_{ac}A^b_{bd}\right)\sigma^{cd} - A^a_{bc}{\partial}_a\sigma^{bc}\right],
\ee
where $\sigma^{ab} = \eta^{ab} -h^{ab}$.  Since the corresponding Feynman diagrams are cubic, the mechanics of graviton perturbation theory are drastically simplified.  From the cubic Feynman rules, it is then straightforward to derive new off-shell recursion relations for graviton amplitudes, in analogy with the Berends-Giele recursion relations for Yang-Mills theory \cite{Berends:1987me,Mafra:2015vca}.  In this cubic representation, the Einstein field equations take the form of two coupled first-order differential equations that are at most quadratic in the fields.  Note that this construction is a field redefinition away from the first-order Palatini formalism \cite{Ferraris1982}, whose cubic structure was emphasized long ago in \Ref{Deser:1969wk}, though not in the context of graviton perturbation theory.

Subsequently, we show how a judicious choice of graviton field basis and gauge fixing yields an especially simple form of the EH action.  In the phenomenologically relevant case of $D=4$ dimensions, we obtain the gauge-fixed action\footnote{Our notational conventions are $T_{(ab)} = T_{ab} + T_{ba}$,  $T_{[ab]} = T_{ab} - T_{ba}$, and $\overset{\leftrightarrow}{\partial}_a = \partial_a - \overset{\leftarrow}{\partial}_a$. Throughout, $\partial_a$ denotes differentiation to the right, while $\overset{\leftarrow}{\partial}_a$ denotes differentiation to the left.}
\be 
S_{\rm EH} + S_{\rm GF} = -\frac{1}{16\pi G}\int \mathrm{d}^4 x \,  K^{ab} \sigma_{ab} , 
\label{eq:SEHsimp}
\ee
where $\sigma_{ab} = \eta_{ab} + h_{ab}  + h^2_{ab} + h^3_{ab}+\cdots$ is the inverse of $\sigma^{ab}$ expressed as a geometric series in the graviton. Here we have defined the kinetic tensor
\be 
 K^{ab} = \frac{1}{2} \partial_{[c} h^{ac} \partial_{d]} h^{bd}  + \frac{1}{4} h^{cd}  \overset{\leftrightarrow}{\partial}_d \partial_c h^{ab} -\frac{1}{4} \eta_{cd} h^{ac} \Box h^{bd},\label{eq:Kintro}
\ee
whose trace $ K^{ab} \eta_{ab}$ corresponds to the graviton kinetic term.  Remarkably, all graviton interaction vertices are given trivially by the kinetic tensor multiplied by powers of the graviton.   This simplicity stands in stark contrast with graviton perturbation theory in the conventional approach, where interaction vertices grow intractably lengthy and complex for increasing powers of the graviton.  

Let us put our results in context with some other recent approaches related to finding simpler ways of calculating quantities in classical and perturbative quantum gravity, as well as applying the techniques of scattering amplitudes to problems in classical and semiclassical gravity \cite{BjerrumBohr:2002kt,Neill:2013wsa}. Indeed, finding ways of simplifying calculations in general relativity is a particularly relevant and pressing problem in light of LIGO's recent detections of gravitational waves \cite{LIGO,Abbott:2016nmj,Abbott:2017vtc}. In particular, the celebrated BCJ double copy \cite{BCJ} relating amplitudes in gauge theory and gravity has been explored in classical contexts \cite{Monteiro:2014cda,Ridgway:2015fdl,Goldberger:2016iau}. In \Ref{twofold}, a field redefinition and gauge fixing of the EH action was found that allowed the Lagrangian to exhibit the twofold Lorentz invariance whose existence was suggested at the level of amplitudes by the double copy; further, the perturbation theory for the action in \Ref{twofold} is simpler than the canonical perturbation theory of the EH action. In this paper, we will make simplicity of the action the goal, independent of consideration of the double copy or twofold Lorentz invariance (though this property will make an appearance in \Sec{sec:enhanced}).

The remainder of this paper is organized as follows.  In \Sec{sec:cubic} we construct a cubic representation of the EH action by integrating in a single auxiliary field.  We then derive Feynman rules and off-shell recursion relations for graviton scattering amplitudes.  Afterward, in \Sec{sec:simp} we derive a further simplified representation of the EH action by exploiting the freedom of gauge fixing.  We then discuss the generalization of our results to curved spacetime in \Sec{sec:curved} and conclude in \Sec{sec:conclusions}.

\section{Cubic Formulation}

\label{sec:cubic}

In this section we reformulate the EH action as a theory of purely cubic interactions.   To do this, we devise a convenient field basis for the graviton in which the action arises from integrating out a single auxiliary field. We derive the corresponding Feynman rules and off-shell recursion relations for graviton scattering amplitudes.  As we will see, the resulting cubic formulation is compact and enjoys an enhanced twofold Lorentz symmetry.

\subsection{Lagrangian}

\subsubsection{Field Basis}

All of our results are derived directly from the EH action in $D$ dimensions,
\be 
S_{\rm EH} = \frac{1}{16\pi G}  \int \mathrm{d}^D x \, \sqrt{-g} \, R ,
\ee
working in mostly-plus signature.
As shown in \Ref{twofold}, the corresponding Lagrangian can be rewritten in the form
\be 
\begin{aligned}
\sqrt{-g}\, R =& \sqrt{-g} \left[ \partial_a g_{ce} \partial_b g^{de} \left( \frac{1}{4} g^{ab} \delta^c_d- \frac{1}{2} g^{cb}\delta^a_d \right) - g^{ab} \partial_a \partial_b (\log \sqrt{-g}) \right]+ \cdots
 \\ =& \sqrt{-g} \bigg[  \partial_a \left(\frac{g_{c e}}{\sqrt{-g}}\right) \partial_b \left(\sqrt{-g} \, g^{de}\right)  \left( \frac{1}{4}  g^{ab} \delta^c_d- \frac{1}{2} g^{cb}\delta^a_d \right)  \\&\qquad\qquad+ \frac{D-2}{4} g^{ab} \partial_a (\log \sqrt{-g}) \partial_b (\log \sqrt{-g}) \bigg]+ \cdots,\label{eq:EHsimp}
\end{aligned}
\ee
where the ellipses denote total derivative contributions that we hereafter neglect.  From the second equality in \Eq{eq:EHsimp}, it is clear that the Lagrangian is naturally a function of the variables\footnote{These are sometimes referred to in the literature as the ``gothic'' variables $\mathfrak{g}^{ab} = \sigma^{ab}$. We will use the $\sigma$ notation for clarity and consistency with \Ref{twofold}.}
\be 
 \sigma_{ab} = \frac{1}{\sqrt{-g}} \, g_{ab} \qquad \textrm{and}\qquad  \sigma^{ab} = \sqrt{-g} \, g^{ab}, \label{eq:sigma_def}
\ee
where by definition the ``lowered'' $\sigma$ fields and ``raised'' $\sigma^{-1}$ fields are inverses of each other, so
\be
\sigma^{ab} \sigma_{bc} = \delta^a_c. \label{eq:identity}
\ee
In terms of the $\sigma$ and $\sigma^{-1}$ fields, the EH action becomes
\be 
S_{\rm EH} = \frac{1}{16\pi G}\int \mathrm{d}^D x \,  {\cal L }_{\rm EH}
\ee
with the associated Lagrangian
\be 
{\cal L }_{\rm EH}  = \partial_a \sigma_{c e} \partial_b \sigma^{de} \left( \frac{1}{4} \sigma^{ab} \delta^c_d- \frac{1}{2} \sigma^{cb}\delta^a_d \right) + \frac{D-2}{4}\sigma^{ab} \omega_a \omega_b,
\label{eq:SEH_sigma}
\ee
where for later convenience we have defined the vector
\be 
\omega_a = \partial_a \log \sqrt{-g}  = \frac{1}{D-2} \sigma_{bc} \partial_a \sigma^{bc},
\ee
which characterizes variations in the volume element.

While the EH action is not conformally invariant, the notion of conformal weight will be a handy bookkeeping tool for terms in the action.
Under a conformal transformation, the metric transforms as
\be 
g_{ab} \rightarrow \Omega^{-2} g_{ab},
\ee
which acts on the natural variables in \Eq{eq:sigma_def} according to
\be 
 \sigma_{ab} \rightarrow  \Omega^{D-2} \sigma_{ab}  \qquad \textrm{and}\qquad \sigma^{ab} \rightarrow  \Omega^{2-D} \sigma^{ab}.
\ee
In particular, the conformal weights are $[g_{ab}  ] =-2$ and $[\sqrt{-g}  ] =-D$ for the metric and volume measure, respectively, and $[{\cal L}_{\rm EH}] = 2-D$ for the Lagrangian, which is consistent with the mass dimension of the gravitational constant, $[G] = 2-D$. In order to abide by the conformal weight counting, the EH Lagrangian must take the schematic form
\be 
{\cal L}_{\rm EH}  \sim \sum_n (\sigma^{-1})^n (\sigma)^{n-1}.
\ee
In other words, every term must carry one more factor of $\sigma^{-1}$ than $\sigma$.   For instance, without any additional manipulation, the EH action in \Eq{eq:SEH_sigma} is of the form ${\cal L}_{\rm EH} \sim (\sigma^{-1})^2 (\sigma) +  (\sigma^{-1})^3 (\sigma)^2  $.  As we will see, the EH action can be rewritten in various forms that are homogeneous in powers of $\sigma$ and $\sigma^{-1}$, i.e., for which ${\cal L}_{\rm EH} \sim (\sigma^{-1})^n (\sigma)^{n-1}$ for a single power $n$.  The EH action has many elegant properties when recast into a homogeneous form.

\subsubsection{Auxiliary Fields}

The conventional approach to graviton perturbation theory entails interaction vertices of arbitrarily high order. That is, the $\mathcal{O}(h^n)$ nonlinearities of the action are present for arbitrarily high $n$.   However, we will now see how this tower of interactions can be resummed by introducing as few as one auxiliary field.  The crux of our construction is to treat $\sigma$ as the fundamental field and generate all factors of $\sigma^{-1}$ by integrating out an auxiliary field (or vice versa with $\sigma$ and $\sigma^{-1}$ swapped).

To be concrete, let us now describe how to recast the EH action in \Eq{eq:SEH_sigma} into the homogeneous form ${\cal L}_{\rm EH} \sim (\sigma^{-1})^3 (\sigma)^2$.  We substitute in \Eq{eq:identity} to transform the $(\sigma^{-1})^2  (\sigma)$ term into a term of the form $(\sigma^{-1})^3 (\sigma)^2$, yielding
\be
\mathcal{L}_{\rm EH} = {\partial}_a \sigma_{bc} \left(-\frac{1}{4} \sigma^{ad} \sigma^{be} \sigma^{cf} + \frac{1}{2}\sigma^{ae} \sigma^{bd}\sigma^{cf} - \frac{1}{4(D-2)} \sigma^{ad} \sigma^{bc}\sigma^{ef}\right){\partial}_d \sigma_{ef}.\label{eq:EHaction3}
\ee
Since \Eq{eq:EHaction3} is a quadratic form in $\sigma$, it is natural to treat this field as fundamental and integrate in an auxiliary field that generates the remaining $\sigma^{-1}$ factors.
By inverting the term in parentheses in \Eq{eq:EHaction3}, we obtain the equivalent action
\be 
\mathcal{L}_{\rm EH} = -A^{abc}\left(\sigma_{ae}\sigma_{bd}-\frac{1}{D-1}\sigma_{ab}\sigma_{de}\right)\sigma_{cf}A^{def} + A^{abc}{\partial}_a \sigma_{bc} ,\label{eq:quintic2}
\ee
where $A^{abc} = A^{acb}$ is an auxiliary field.   Note that \Eq{eq:quintic2} is fully equivalent to the EH action, albeit with interactions that truncate at quintic order.  

This procedure generalizes in the obvious way.  By inserting the Kronecker delta function in \Eq{eq:identity} into \Eq{eq:EHaction3} in various ways, we can rearrange the Lagrangian into a form with interaction vertices that truncate at any arbitrary but finite order. For example, from \Eq{eq:quintic2} one can derive an alternative quintic action in terms of $\sigma^{ab}$ rather than $\sigma_{ab}$, plus an auxiliary field with all lowered indices. As we will soon see, the minimal construction of this type results in a cubic Lagrangian.

Returning to \Eq{eq:quintic2}, we derive the equation of motion for $A^{abc}$,
\be
A^{abc} = \sigma^{bd}\sigma^{ce}\Gamma^a_{de}- \frac{1}{2}\sigma^{a(b}\sigma^{c)d}\Gamma^e_{de}, 
\ee
where $\Gamma^a_{bc}$ is the Christoffel symbol written as an implicit function of the metric in terms of $\sigma$ and $\sigma^{-1}$ through  \Eq{eq:sigma_def}.  From the above relation, it is natural to define a new auxiliary field with the same index structure as the Christoffel symbol,
\be  
A^a_{bc} = A^{ade}\sigma_{bd}\sigma_{ce},\label{eq:ChristoffelA}
\ee
so the action in \Eq{eq:quintic2} takes an even simpler form,
\be 
\mathcal{L}_{\rm EH} =  -\left(A^a_{bc} A^b_{ad} - \frac{1}{D-1} A^a_{ac}A^b_{bd}\right)\sigma^{cd} + A^a_{bc}{\partial}_a\sigma^{bc}.\label{eq:cubicA1}
\ee
In this basis, the natural field variable is $\sigma^{-1}$ and integrating out $A$ generates all factors of $\sigma$. The Lagrangian in \Eq{eq:cubicA1} is a primary result of this paper: a cubic representation of the EH action in terms of the graviton and a single auxiliary field.

Since \Eq{eq:cubicA1} is equivalent to the EH action, the associated equations of motion are equivalent to the Einstein field equations.  The equation of motion for the graviton field $\sigma^{ab}$ is
\be 
\frac{\delta \mathcal{L}_{\rm EH}}{\delta \sigma^{ab}}  = -A^c_{da}A^d_{cb} + \frac{1}{D-1}A^c_{ca}A^d_{db} - \partial_c A^c_{ab} = 0.\label{eq:eomsigma}
\ee
Note that the left-hand side is equal to the Ricci tensor, ${\delta \mathcal{L}_{\rm EH}}/{\delta \sigma^{ab}} = R_{ab}$, which follows from the Jacobian relating $g^{ab}$ and $\sigma^{ab}$, as derived in \Ref{twofold}. Meanwhile, the equation of motion for the auxiliary field is 
\be 
\frac{\delta \mathcal{L}_{\rm EH}}{\delta A^a_{bc}} = -\left(A^{(b}_{ad} - \frac{1}{D-1} A^e_{ed} \delta^{(b}_a\right) \sigma_{\vphantom{}}^{c)d} + \partial_a \sigma^{bc} = 0.\label{eq:eomA}
\ee
The two coupled first-order differential equations in Eqs.~\eqref{eq:eomsigma} and \eqref{eq:eomA} are equivalent to the Einstein field equations.

Let us now comment on one final cubic representation of the action.  After some rearrangement, \Eq{eq:eomA} can be written as 
\be
A^a_{bc} = \Gamma^a_{bc}  - \frac{1}{2}\delta^a_{(b} \Gamma^d_{c)d}.\label{eq:eomChristoffel}
\ee 
Motivated by the link between the auxiliary field and the Christoffel symbol, we go to a field basis in which the auxiliary field is literally equal to the Levi-Civita connection on shell, so
\be
B^a_{bc}=A^a_{bc}-\frac{1}{D-1}\delta^a_{(b} A^d_{c)d}
\ee
and the action becomes
\be 
\mathcal{L}_{\rm EH} = -B^a_{bc}\left(\delta^e_a\sigma^{cf}-\delta^c_a\sigma^{ef}\right)B^b_{ef}-B^a_{bc}{\partial}_a\sigma^{bc}+B^c_{bc}{\partial}_a\sigma^{ab}.
\label{eq:cubicX}
\ee 
By construction, the equations of motion set $B^a_{bc}=\Gamma^a_{bc}$ on shell.  Partially integrating \Eq{eq:cubicX}, we obtain yet another alternative form of the EH action, 
\be 
S_{\rm EH} = \frac{1}{16\pi G} \int  \mathrm{d}^{D}x \, \sqrt{-g} \, g^{ab}\left(\partial_{c}B_{ba}^{c}-\partial_{b}B_{ca}^{c}+B_{cd}^{c}B_{ba}^{d}-B_{bd}^{c}B_{ca}^{d}\right),
\ee
plugging in $\sqrt{-g} \, g^{ab}= \sigma^{ab}$ from the definition in \Eq{eq:sigma_def}.   Substituting $\Gamma^a_{bc}$ for $B^a_{bc}$ in the expression in parentheses yields the Ricci tensor $R_{ab}$ written in terms of Christoffel symbols. The action in \Eq{eq:cubicX} is closely related to the Palatini formalism \cite{Ferraris1982} in which one takes the EH action and treats the connection as a priori independent of the metric; see also Refs.~\cite{Deser:1969wk,Parattu:2013gwa}.

\subsection{Perturbation Theory}\label{sec:recursion}

The Lagrangians in Eqs.~\eqref{eq:quintic2}, \eqref{eq:cubicA1}, and \eqref{eq:cubicX} treat either $\sigma$ or $\sigma^{-1}$ as the fundamental fields. However, since we have made no assumptions about the size of the field values, these actions apply for arbitrarily large deviations away from flat space.  This is the case even though we have chosen to write these Lagrangians purely in terms of partial rather than covariant derivatives.

On the other hand, it is still of practical interest to study gravity perturbatively in powers of graviton fluctuations $h_{ab}$ about a flat background in Cartesian coordinates, $\eta_{ab} = \mathrm{diag}(-1,1,\ldots,1)$.  For Lagrangians in which the fundamental field is $\sigma_{ab}$, we define $\sigma_{ab} = \eta_{ab} + h_{ab}$, in which case $\sigma^{ab}$ is a geometric series in the graviton.  Meanwhile, for those Lagrangians in which the fundamental field is $\sigma^{ab}$, we use a different but physically equivalent field basis $\sigma^{ab} = \eta^{ab} - h^{ab}$, for which $\sigma_{ab}$ is a geometric series. Note that these two uses of $h_{ab}$ are inequivalent, but are related to each other by a field redefinition (and similarly are related to the graviton field in canonical perturbation theory via a different field redefinition).

Though elegant, the action in \Eq{eq:cubicA1} is not yet in a form appropriate for perturbation theory, since there is explicit mixing between the graviton and the auxiliary field.  In this section, we will show how to unmix these states and derive the propagators and Feynman vertices for the corresponding graviton perturbation theory. 
To eliminate the mixing between the gravition $h_{ab}$ and the auxiliary field $A^a_{bc}$, we apply the field shift
\be
A^a_{bc}\rightarrow A^a_{bc} - \frac{1}{2}\left(\partial_b h^a_{\;\;c} + \partial_c h^a_{\;\;b} - \partial^a h_{bc} + \frac{1}{D-2}\eta_{bc}\partial^a h^d_{\;\;d}\right),\label{eq:unmix}
\ee
where indices on $h_{ab}$ and $\partial_a$ are raised and lowered using the background metric $\eta_{ab}$. 

\begin{figure}[t]
  \begin{center}
    \includegraphics[width=\textwidth]{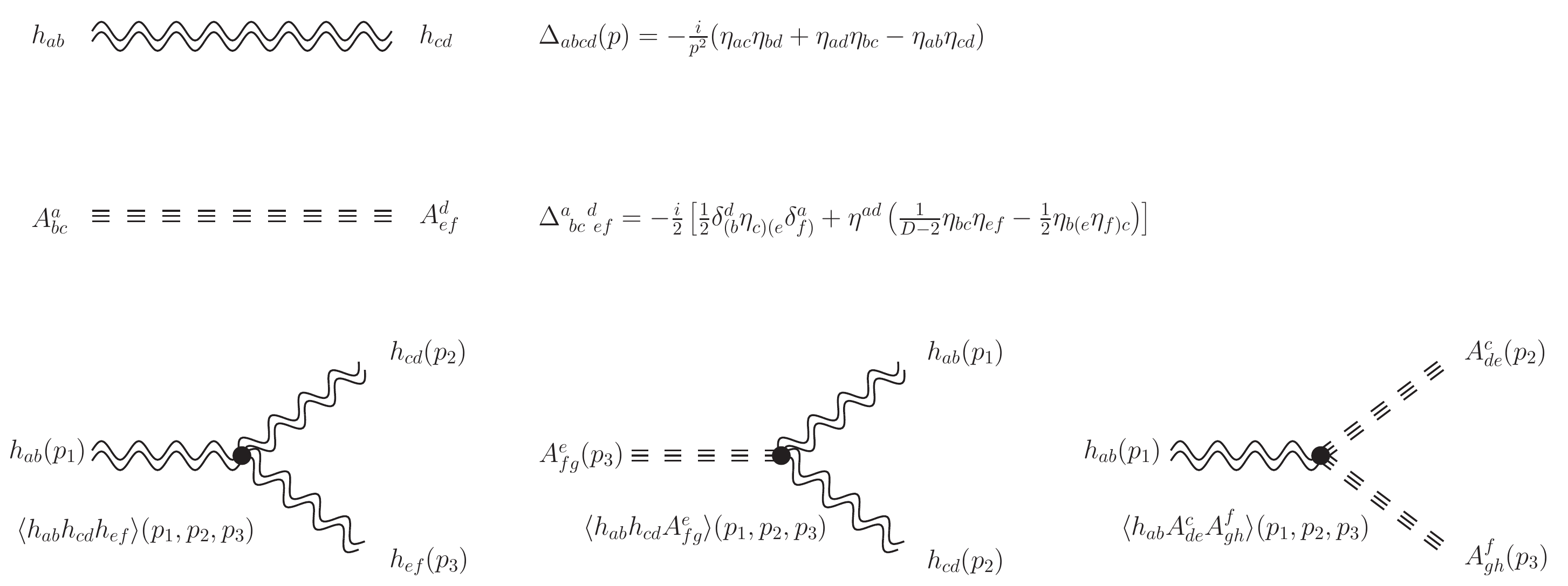}
  \end{center}
  \caption{Summary of Feynman propagators and vertices for the cubic gravity action in \Eq{eq:actionunmixed}.}
\label{fig:Ahtheory}
\end{figure}

After diagonalizing the quadratic term in \Eq{eq:cubicA1}, we add the gauge-fixing term  
\be 
\mathcal{L}_{\rm GF} = -\frac{1}{2}\partial_a h^{ac} \partial^b h_{bc} = -\frac{1}{2} \eta_{cd} \partial_a (\sqrt{-g} g^{ac}) \partial_b (\sqrt{-g} g^{bd}),
\ee
so that the graviton propagator is well defined. This gauge choice coincides with harmonic (de~Donder) gauge for $\sigma^{ab}$, i.e., the requirement $\partial_a h^{ab} = 0$ for the trace-reversed field $h_{ab} - \frac{1}{2}\eta_{ab}h^c_{\;\;c}$. Upon gauge fixing, the Lagrangian\footnote{For notational convenience, we suppress the $16\pi G$ normalization of the action in our discussion of the Feynman rules.  To convert to the canonically normalized scattering amplitudes, simply multiply the amplitude computed using our Feynman rules by a factor of $1/16\pi G$, together with a factor of $\sqrt{32 \pi G}$ for each external graviton.}  becomes
\be 
\mathcal{L} = \mathcal{L}_{\rm EH} + \mathcal{L}_\mathrm{GF} =  {\cal L}_{hh} + {\cal L}_{AA} + {\cal L}_{hhh}  + {\cal L}_{hhA}+  {\cal L}_{hAA},\label{eq:actionunmixed}
\ee
where the quadratic terms are
\be 
\begin{aligned}
{\cal L}_{hh}  &= \frac{1}{4}\left(h_{ab} \Box h^{ab} - \frac{1}{D-2}[h]\Box[h]\right)\\
{\cal L}_{AA}  &= - \left(A^a_{bc} A^b_{ad} - \frac{1}{D-1} A^a_{ac}A^b_{bd} \right)\eta^{cd}
\end{aligned}
\ee
and the cubic terms are \be 
\begin{aligned}
{\cal L}_{hhh} &= \frac{1}{4}h^{ab}\left[\partial_a h_{cd} \partial_b h^{cd} + 2\partial_{[c} h_{d]b} \partial^d h_a^{\;\;c} + \frac{1}{D-2}\left(2 \partial_c h_{ab}\partial^c [h] - \partial_a [h] \partial_b [h]\right) \right]\\
{\cal L}_{hhA}  & = h^{ab} \left[A^c_{ad}(\partial^d h_{bc}  - \partial^{\vphantom{}}_{(b} h_{c)}^{\;\;d}) -\frac{1}{D-2}(\eta_{ad}A^d_{bc} \partial^c [h] - A^c_{ca} \partial_b [h])\right]\\
{\cal L}_{hAA}  & = h^{ab} \left(A^c_{ad} A^d_{bc} -\frac{1}{D-1}A^c_{ac} A^d_{bd}\right),
\end{aligned}
\ee
where $[h] = h^a_{\;\;a}$. 
In this gauge, the graviton propagator takes the simple $D$-independent form 
\be 
\Delta_{abcd}
= -\frac{i}{p^2}(\eta_{ac} \eta_{bd} + \eta_{ad} \eta_{bc} - \eta_{ab}\eta_{cd}),\label{eq:gravprop}
\ee
corresponding to propagation from $h_{ab}$ to $h_{cd}$.  The auxiliary field propagator takes the form
\be 
\Delta^{a\,\;\,\;d}_{\,\;bc\,\;ef} 
= -\frac{i}{2} \left[\frac{1}{2}\delta_{(b}^d \eta^{\vphantom{}}_{c)(e}\delta_{f)}^a+\eta^{ad}\left(\frac{1}{D-2}\eta_{bc}\eta_{ef} -\frac{1}{2}\eta_{b(e}\eta_{f)c}\right)\right],
\ee
corresponding to propagation from $A^a_{bc}$ to $A^d_{ef}$.  Meanwhile, the interaction vertices are 
\be 
\begin{aligned}
\langle h_{ab} h_{cd}h_{ef}\rangle (p_1,p_2,p_3)&= \frac{i}{4}\Big\{ \Big[ \frac{1}{2}(\eta_{a(c}\eta_{d)(e}\eta_{f)b} +\eta_{b(c}\eta_{d)(e}\eta_{f)a})(p_1 p_2)\\&\qquad -\frac{1}{D-2}  (\eta_{ab}\eta_{c(e}\eta_{f)d} + \eta_{cd}\eta_{a(e}\eta_{f)b})(p_1 p_2) \\&\qquad  +\left(\frac{1}{D-2}\eta_{ab}\eta_{cd}-\frac{1}{2}\eta_{a(c}\eta_{d)b}\right)p_{1(e}p_{2f)} -\frac{1}{2}p_{2(a} \eta_{b)(e} \eta_{f)(d} p_{1c)} \Big] \\&\qquad + \Big[\stacklines{p_2\leftrightarrow p_3}{cd\leftrightarrow ef } \Big] + \Big[\stacklines{p_1 \leftrightarrow p_3}{ab\leftrightarrow ef}\Big]\Big\} \\
  \langle h_{ab}h_{cd} A^{e}_{fg}\rangle(p_1,p_2,p_3) &= \frac{1}{4} \Big\{ \Big[\frac{1}{2}\delta_{(a}^e\left( \eta^{\vphantom{}}_{b)(f}\eta^{\vphantom{}}_{g)(c}p^{\vphantom{}}_{1d)} - \eta^{\vphantom{}}_{b)(c}\eta^{\vphantom{}}_{d)(f}p^{\vphantom{}}_{1g)}\right) \\ &\qquad + \frac{1}{D-2} \eta^{\vphantom{}}_{ab} \left(p^{\vphantom{}}_{1(f}  \eta^{\vphantom{}}_{g)(c} \delta_{d)}^e -p^{\vphantom{}}_{1(c}  \eta^{\vphantom{}}_{d)(f} \delta_{g)}^e\right) \Big] + \Big[ \stacklines{p_1\leftrightarrow p_2}{ab\leftrightarrow cd} \Big] \Big\} \\& \qquad -\frac{1}{8}p_3^e \left(\eta_{f(a}\eta_{b)(c}\eta_{d)g} +\eta_{g(a}\eta_{b)(c}\eta_{d)f}\right)\\
\langle h_{ab}  A^{c}_{de}  A^{f}_{gh}\rangle (p_1,p_2,p_3) &= \frac{i}{4}\left(\delta_{(g}^c\eta^{\vphantom{}}_{h)(a}\eta^{\vphantom{}}_{b)(d}\delta_{e)}^f - \frac{1}{D-1}\delta_{(g}^f\eta^{\vphantom{}}_{h)(a}\eta^{\vphantom{}}_{b)(d}\delta_{e)}^c \right).
\end{aligned}
\ee
The above Feynman rules are summarized in \Fig{fig:Ahtheory}.

\subsection{Recursion Relations}\label{sec:recursionrelations}

Since the Lagrangian in \Eq{eq:cubicA1} is comprised of purely cubic interactions, we can derive explicit off-shell recursion relations for tree-level graviton scattering amplitudes in analogy with the Berends-Giele recursion relations for Yang-Mills theory.  In fact, these gravity recursion relations are in a sense simpler than for Yang-Mills theory, as the action in \Eq{eq:cubicA1} has no quartic interactions. 

To begin, let us define the off-shell graviton current $J_{ab}(p_\alpha)$, corresponding to an insertion of a graviton field $h_{ab}$ branching off into a set $\alpha$ of on-shell gravitons, and the off-shell auxiliary field current $J^a_{bc}(p_\alpha)$, corresponding to an insertion of the auxiliary field  $A^a_{bc}$ branching off into a set $\alpha$ of on-shell gravitons.   Here the dependence on momentum flowing through the current, $p_\alpha = \sum_{i \in \alpha} p_i$, is shown explicitly.  The currents are also implicit functions of the momenta and polarization tensors of the remaining on-shell external states.    We adopt a convention in which the on-shell gravitons are incoming and the off-shell leg is outgoing, while all gravitons are incoming for the vertices.  The currents are equal to
\be 
\begin{aligned}
J_{ab}(p_\alpha) &= i \Delta_{abcd}(p_\alpha)  M^{cd}(p_\alpha) \\
J^{a}_{bc}(p_\alpha) &= i  \Delta^{a\,\;\,\;d}_{\,\;bc\,\;ef}  (p_\alpha) M^{ef}_d(p_\alpha),
\end{aligned}
\ee
where $M_{ab}$ and $M^a_{bc}$ are semi-on-shell amplitudes with all legs on-shell except for one leg with momentum $p_\alpha$  corresponding to an off-shell graviton or auxiliary field, respectively.

The graviton current satisfies the recursion relations
\be
\begin{aligned}
J_{ab}(p_\alpha) = \Delta_{abcd}(p_\alpha) \sum_{\substack{\alpha_1 \cup \alpha_2 = \alpha}}  \bigg[ &+  \frac{1}{2} \langle h^{cd} h^{ef} h^{gh} \rangle(-p_{\alpha},p_{\alpha_1},p_{\alpha_2}) J_{ef}(p_{\alpha_1})J_{gh}(p_{\alpha_2}) \\   
& + \langle h^{cd}h^{ef} A_{g}^{hi}\rangle(-p_{\alpha},p_{\alpha_1},p_{\alpha_2}) J_{ef}(p_{\alpha_1})J^g_{hi}(p_{\alpha_2})
 \\  & +  \frac{1}{2}    \langle h^{cd}A_e^{fg} A_{h}^{ij} \rangle (-p_{\alpha},p_{\alpha_1},p_{\alpha_2}) {J}^e_{fg}(p_{\alpha_1}){J}^h_{ij}(p_{\alpha_2})\bigg], \label{eq:recursion1}
 \end{aligned}
 \ee
 while the auxiliary field current satisfies
\be 
\begin{aligned}
J^a_{bc} (p_\alpha)= \Delta^{a\,\;\,\;d}_{\,\;bc\,\;ef}(p_\alpha) \sum_{\alpha_1 \cup \alpha_2 = \alpha} \bigg[  & + \frac{1}{2}\langle h^{gh}h^{ij} A_{d}^{ef} \rangle (p_{\alpha_1},p_{\alpha_2},-p_{\alpha})  J_{gh}(p_{\alpha_1}) J_{ij}(p_{\alpha_2})\\ &+    \langle h^{gh}A_i^{jk} A_{d}^{ef} \rangle (p_{\alpha_1},p_{\alpha_2},-p_{\alpha})J_{gh}(p_{\alpha_1}) J^i_{jk}(p_{\alpha_2}) \bigg],\label{eq:recursion2}
\end{aligned}
\ee
where each sum runs over all partitions of the set $\alpha$ of on-shell graviton labels into distinct subsets $\alpha_1$ and $\alpha_2$. 

As with the Berends-Giele recursion relations, the above equations are to be solved iteratively, order by order in the number of external on-shell gravitons.  The initialization step of the recursion relations involves just a single on-shell graviton, where $J_{ab}(p) = \epsilon_{ab}$ and $J^a_{bc} (p) =0 $.  The latter vanishes because we are interested only in currents with gravitons as on-shell external states and because we have used the transformation in \Eq{eq:unmix} to obtain the action in \Eq{eq:actionunmixed} in which the graviton and auxiliary field do not mix.
 Using the recursion relations in Eqs.~\eqref{eq:recursion1} and \eqref{eq:recursion2}, we have calculated the off-shell graviton current up to fourth order in on-shell gravitons, obtaining the correct three-particle, four-particle, and five-particle amplitudes.
 
 \subsection{Enhanced Symmetries}\label{sec:enhanced}

The study of graviton scattering amplitudes has revealed a number of noteworthy surprises, including enhanced cancellations in supergravity theories \cite{Bern:2014sna,Bern:2012gh,Bern:2012cd} and the so-called ``bonus relations'' that arise in the BCFW recursion relations \cite{Spradlin:2008bu,ArkaniHamed:2008gz}.   Another miraculous result is the celebrated ``double copy'' construction relating graviton scattering amplitudes to the squares of gluon amplitudes, e.g., via the KLT \cite{KLT} and BCJ \cite{BCJ} relations. In the former representation, graviton scattering amplitudes are expressed as products of Lorentz invariant gluon amplitudes, suggesting a hidden twofold Lorentz invariance within gravity. In \Ref{twofold} it was shown that with a careful choice of field basis and gauge fixing one obtains a form of the EH action that exhibits this symmetry explicitly.  At the level of the action, twofold Lorentz invariance is manifested as a consistent labeling of all indices as one of two types (e.g., barred and unbarred indices as in $h_{a \bar b}$), which are separately contracted \cite{Bern:1999ji,Hohm:2011dz}.

Though not by design, the cubic action in \Eq{eq:cubicA1} automatically exhibits a twofold Lorentz symmetry in the spirit of the construction in \Ref{twofold}.   
In particular, one can write \Eq{eq:cubicA1} as
\be 
\begin{aligned}
{\cal L}_{\rm EH} =& -\frac{1}{2}\left(A^{\bar a}_{c \bar b} A^{\bar b}_{d \bar a} - \frac{1}{D-1} A^{\bar a}_{c \bar a}A^{\bar b}_{d \bar b}\right)\eta^{cd}-\frac{1}{2}\left(A^{a}_{b \bar c} A^{b}_{a \bar d} - \frac{1}{D-1} A^{a}_{a \bar c}A^{b}_{b \bar d}\right)\eta^{\bar c \bar d} \\&+ \frac{1}{2}\left(A^a_{c \bar b} A^{\bar b}_{a \bar d} - \frac{1}{D-1} A^{\bar a}_{c \bar a}A^b_{b \bar d}\right)h^{c \bar d}+ \frac{1}{2}\left(A^{\bar a}_{ b \bar c} A^b_{d \bar a} - \frac{1}{D-1} A^{a}_{a \bar c}A^{\bar b}_{d \bar b}\right)h^{d \bar c} \\&- \frac{1}{2}\left(A^a_{b \bar c}{\partial}_a h^{b \bar c}+A^{\bar a}_{b \bar c}{\partial}_{\bar a} h^{b \bar c}\right),
\label{eq:factorized}
\end{aligned}
\ee
where as before we have expanded in $\sigma^{ab} = \eta^{ab} - h^{ab}$ before promoting the graviton to a general tensor $h_{a \bar b}$ and the auxiliary field to a pair of fields $A^a_{b\bar c}$ and $A^{\bar a}_{b \bar c}$.
 Derivatives can carry either unbarred or barred indices, $\partial_a$ or $\partial_{\bar a}$, while the metric enters either as $\eta_{ab}$ or $\eta_{\bar a \bar b}$. The action in \Eq{eq:factorized} is explicitly invariant under an $SO(D-1,1)\times \overline{SO}(D-1,1)$ twofold Lorentz symmetry that acts separately on barred and unbarred indices. As discussed in \Ref{twofold}, this symmetry can be made manifest in the Lagrangian by introducing an auxiliary extra set of spacetime dimensions.

That there exists a simple cubic formulation of the EH action with twofold Lorentz invariance is particularly enticing given the BCJ prescription in which graviton amplitudes are obtained by squaring the numerators of gluon amplitudes that are expressed in a particular cubic form.  A BCJ-compliant action for gravity is conceivable, since BCJ duality has already been demonstrated as a manifest symmetry of a particular cubic representation of the nonlinear sigma model \cite{Cheung:2016prv}.  Unfortunately, when the graviton and auxiliary fields are unmixed, the twofold Lorentz invariance is no longer manifest.

\section{Simplified Formulation}\label{sec:simp}

Thus far we have only exploited the freedom of choosing a field basis to simplify the Lagrangian.  However, gauge fixing also offers enormous leeway in reformulating the action.\footnote{Though we introduced a simple gauge-fixing term in \Sec{sec:recursion}, we did not use this freedom to make the perturbation theory of the pure gravity action as simple as possible.}  As we will see, with an appropriate nonlinear gauge fixing it is possible to eliminate the $(\sigma^{-1})^3 (\sigma)^2$ term in \Eq{eq:SEH_sigma} in order to write the action in the homogeneous form  ${\cal L}_{\rm EH} \sim (\sigma^{-1})^2 (\sigma)$.   
  The resulting action is strikingly simple, allowing for a closed-form expression for graviton interaction vertices at arbitrarily high order.

\subsection{Lagrangian}

\subsubsection{Gauge Fixing}

We define the full gauge-fixed action to be
\be 
S = S_{\rm EH} +S_{\rm GF},
\ee
where the gauge-fixing term is chosen to be
\be 
S_{\rm GF} = - \frac{1}{16\pi G}\int \mathrm{d}^D x \, \frac{D-2}{4}\sigma^{ab} (\omega_a - \tau_a)(\omega_b - \tau_b)  \label{eq:SGF}
\ee
for some vector $\tau_a$.
With the benefit of hindsight, we make the special choice
\be 
\tau_a = z \sigma_{ab} \partial_c \sigma^{bc} \label{eq:tau}
\ee
for some constant $z$. The gauge-fixing term in \Eq{eq:SGF} corresponds to the gauge condition $\omega_a - \tau_a = 0$. In terms of coordinates $x^a$, which are treated as $D$ real scalar functions on the spacetime manifold, the gauge condition is equivalent to a condition on the coordinates,
\be
\nabla_a \nabla_b x^a =-zg_{ab}\nabla_c \nabla^c x^a,
\ee
that is, $\left(\delta^c_a \delta^d_b + z g_{ab}g^{cd}\right)\nabla_c \nabla_d x^a = 0$, where $\nabla_a$ is the covariant derivative defined with respect to the full metric $g_{ab}$. Our gauge condition for general $z$ is thus a hybrid of the harmonic and unimodular gauge conditions. To derive this coordinate condition, we used the geometric identities $ \Gamma^a_{ba} = \partial_b \log \sqrt{-g}$ and $g^{bc}\Gamma^a_{bc}=-\partial_b (\sqrt{-g}g^{ab})/\sqrt{-g}$.

Reshuffling terms and dropping total derivatives, we obtain the full gauge-fixed action
\be 
S = \frac{1}{16\pi G}\int \mathrm{d}^D x \,  {\cal L },
\ee
where the Lagrangian is given by
\be 
\begin{aligned}
\mathcal{L}  &= -  \left[ \frac{z^2}{4} (D-2) \partial_c \sigma^{ac} \partial_d \sigma^{bd} - \frac{1}{2} \partial_d \sigma^{ac} \partial_c  \sigma^{bd} + \frac{1}{4}\left(1-2z\right) \partial_d \sigma^{cd}  \partial_c \sigma^{ab}  + \frac{1}{4} \sigma^{cd} \partial_c \partial_d \sigma^{ab} \right]\sigma_{ab} .  \label{eq:Sfull}
\end{aligned}
\ee
As advertised, every term in \Eq{eq:Sfull} has two $\sigma^{-1}$ fields and one $\sigma$ field.    To turn this property to our advantage, we again use a field basis in which the graviton perturbations $h_{ab}$ enter as 
\be
\sigma^{ab} = \eta^{ab} - h^{ab}\qquad  \textrm{and} \qquad
\sigma_{ab} = \eta_{ab} + h_{ab}  + h^2_{ab} + h^3_{ab}+  \cdots = \left(\frac{1}{1-h}\right)_{ab},
\ee
where as before $\sigma_{ab}$ is simply a geometric series in the graviton field.   Rearranging terms via integration by parts, we write \Eq{eq:Sfull} as
\be 
\mathcal{L} = -K^{ab} \sigma_{ab},  \label{eq:SEH_K}
\ee
where the kinetic tensor $K^{ab}$ is a two-derivative quadratic form in the graviton,
\be 
\begin{aligned}
 K^{ab} =& + \frac{z^2}{4} (D-2) \partial_c h^{ac} \partial_d h^{bd} -  \frac{1}{2}\partial_d h^{ac} \partial_c  h^{bd} + \frac{1}{4}\left(1-2z\right) \partial_d h^{cd}  \partial_c h^{ab} \\ & + \frac{1}{4} h^{cd} \partial_c \partial_d h^{ab} -\frac{1}{4} \eta_{cd} h^{ac} \Box h^{bd} . \label{eq:Kgen}
 \end{aligned}
 \ee
The graviton kinetic term is given by $-K^{ab}\eta_{ab}$, while all higher-order interactions are simply related to this term by trivial powers of $h_{ab}$.  That is, the graviton interaction is of fixed length and complexity to arbitrarily high order in the graviton.   This contrasts sharply with the conventional picture of graviton perturbation theory, where tremendous effort is required to compute the interaction vertex at any given order.
Restricting to $D=4$ and setting $z=1$, we obtain another primary result of this paper, given by the action defined in Eqs.~\eqref{eq:SEHsimp} and \eqref{eq:Kintro}.

\subsubsection{Auxiliary Fields}

Just as in \Sec{sec:cubic}, the EH action in \Eq{eq:SEH_K} can be reformulated as a simple cubic theory of the graviton plus auxiliary fields.  That this is possible should be unsurprising since the theory, in terms of graviton perturbations, has the structure of a simple geometric series.  Specifically, we find that \Eq{eq:SEH_K} is generated by the cubic action
\be
\mathcal{L} = A_{ab}B^{ab} - K^a_{\;\;a} + (K^{ab} - A^{cb}h_c^{\;\;a})(B_{ab} - h_{ab}),\label{eq:cubic}
\ee
where $A_{ab}$ and $B_{ab}$ are general two-index fields. On shell, one has $A_c^{\;\;a} = -K^{ab} \sigma_{bc}$, which after plugging back into \Eq{eq:cubic} yields \Eq{eq:SEH_K}.

We emphasize here that all of the nontrivial derivative structure of gravity is encoded in the kinetic tensor $K^{ab}$.  In particular, the kinetic tensor shoulders triple duty, forming the basis of the graviton kinetic term, the $h^3$ interactions, and the $h^2 B$ interactions.   The remaining terms---the $AB$ quadratic term and the  $h^2 A$ and $h AB$ interactions---all have trivial index structure.

\subsection{Perturbation Theory}

In this section we derive the Feynman rules for the action in \Eq{eq:SEH_K}.
As we will see, the interaction vertices are extremely simple. Let us first compute the propagator in our chosen field basis and gauge fixing. Following \Ref{Brandt}, a general graviton propagator can be expanded as
\be 
\Delta_{abcd}  = -\frac{i}{p^2} \sum_{n=1}^5 c^{(n)} T^{(n)}_{abcd},
\ee
where the basis tensors are
\be 
\begin{aligned}
T^{(1)}_{abcd} &= \eta_{ac}\eta_{bd}+ \eta_{ad}\eta_{bc}\\
T^{(2)}_{abcd} &= \eta_{ab}\eta_{cd}   \\
T^{(3)}_{abcd} &= \frac{1}{p^2}\left(p_{a}p_{c}\eta_{bd} + p_{a}p_{d}\eta_{bc}+p_{b}p_{d} g_{ac} + p_{b}p_{c} g_{ad}\right) \\
T^{(4)}_{abcd} &= \frac{1}{p^2}\left(p_{a}p_{b}\eta_{cd} + p_{c}p_{d}\eta_{ab}\right) \\
T^{(5)}_{abcd} &= \frac{1}{p^4} p_{a}p_{b} p_{c}p_{d}. 
\end{aligned}
\ee
Inverting the kinetic term associated with \Eq{eq:Kgen}, we obtain the coefficients for the tensor structures in the graviton propagator: 
\be 
\begin{aligned}
c^{(1)}&= 1 \\
c^{(2)}&= -\frac{2z^2}{(1+z)^2}\\
c^{(3)}&= \frac{2}{z^2(D-2)}-1\\
c^{(4)}&= \frac{2z(z-1)}{(1+z)^2}\\
c^{(5)}&= -\frac{8}{z^2(D-2)}+\frac{8z}{(1+z)^2}. 
\end{aligned}
\ee
Considering $D=4$ and choosing $z=1$ for the gauge fixing, we find that the propagator takes a particularly simple form,
\be 
\Delta_{abcd}(p) = -\frac{i}{p^2}\left(\eta_{ac}\eta_{bd}+ \eta_{ad}\eta_{bc} -\frac{1}{2} \eta_{ab}\eta_{cd}-\frac{2}{p^4} p_{a}p_{b} p_{c}p_{d} \right).\label{eq:4Dprop}
\ee
The geometric series form of our gravity action in \Eq{eq:SEH_K} means that the graviton interactions have precisely the same structure as the kinetic term. As a result, we can write down an analytic formula for all Feynman vertices at any order. We first define $K^{ab}=h^{cd}K_{cdef}^{ab}h^{ef}$, where
\be
K^{ab}_{cdef} = \delta^a_c \delta^b_e \left[\frac{z^2}{4}(D-2)\overset{\leftarrow}{\partial}_d \partial_f - \frac{1}{2} \overset{\leftarrow}{\partial}_f \partial_d - \frac{1}{4} \eta_{df} \Box\right] + \delta^a_e \delta^b_f \left[\frac{1}{4}(1-2z)\overset{\leftarrow}{\partial}_d \partial_c + \frac{1}{4}\partial_d \partial_c \right].\label{eq:Kbig}
\ee
The $\mathcal{O}(h^n)$ term in the action \eqref{eq:SEH_K} is $-h^{cd}K_{cdef}^{ab}h^{ef}h_{ab}^{n-2}$, where $h^n_{ab} = h_a^{\;\;c_1} h_{c_1}^{\;\;c_2} \cdots h_{c_n b}$. Thus, the corresponding $n$-point Feynman vertex $\langle h^{a_1 b_1} \cdots h^{a_n b_n} \rangle (p_1,\ldots, p_n)$ is
\be
\begin{aligned}
& \bigg[-\frac{i}{2^n} \sum_{\sigma\in S_\alpha} \eta^{a_{\sigma_3}(b_{\sigma_4}}\eta^{a_{\sigma_4})(b_{\sigma_5}} \cdots \eta^{a_{\sigma_{n-1}})(b_{\sigma_n}} K^{a_{\sigma_n}) b_{\sigma_3}( a_{\sigma_1} b_{\sigma_1})( a_{\sigma_2} b_{\sigma_2})} (p_{\sigma_1},p_{\sigma_2})\bigg] + \bigg[a_{\sigma_3}\leftrightarrow b_{\sigma_3}\bigg],\label{eq:npointvertex}
 \end{aligned}
\ee
where $K^{ab}_{cdef}(p,q)$ is the momentum-space version of \Eq{eq:Kbig} obtained by sending $\overset{\leftarrow}{\partial}_a$ and ${\partial}_a$ to $i p_a$ and $i q_a$, respectively, and where we have raised all indices via $K^{abcdef}(p,q) = \eta^{cg}\eta^{dh}\eta^{ei}\eta^{fj} K^{ab}_{ghij}(p,q)$. The sum in \Eq{eq:npointvertex} runs over each element $\sigma = \{\sigma_1,\ldots,\sigma_n\}$ of the symmetric group $S_\alpha$ of permutations on the set $\alpha = \{1,\ldots,n\}$ of the $n$ legs.  For the special case of the three-particle vertex, we obtain
\be 
\begin{aligned}
 \langle h^{ab} h^{cd} h^{ef} \rangle(p_1,p_2,p_3)   =  &-\frac{i}{8}  [K^{(ab)(cd)(ef)}(p_2,p_3)+K^{(ab)(ef)(cd)}(p_3,p_2)\\& \;\;\;\, +K^{(cd)(ab)(ef)}(p_1,p_3) + K^{(cd)(ef)(ab)}(p_3,p_1)\\& \;\;\;\, +K^{(ef)(ab)(cd)}(p_1,p_2)+K^{(ef)(cd)(ab)}(p_2,p_1)],
\end{aligned}
\ee
which is in agreement with the known three-particle amplitude.

\section{Curved Spacetime}

\label{sec:curved}

The previous sections were dedicated to constructing graviton perturbation theory about a flat background.  However,  it is straightforward to extend these results to perturbations about a general curved background spacetime with metric  $\tilde{g}_{ab}$.  To accomplish this, we first define the curved spacetime analogues of the field variables in \Eq{eq:sigma_def},
\be 
\sigma_{ab} = \frac{\sqrt{-\tilde{g}}}{\sqrt{-g}} g_{ab} \qquad \mathrm{and}\qquad  \sigma^{ab} = \frac{\sqrt{-g}}{\sqrt{-\tilde{g}}} g^{ab},\label{eq:sigmacurved}
\ee
which we employ for the remainder of this section.
As shown in \Ref{twofold}, the curved spacetime generalization of \Eq{eq:EHsimp} is
\be
S_{\rm EH} = \frac{1}{16\pi G}\int \mathrm{d}^D x \, \sqrt{-\tilde g}\left[\tilde{\nabla}_a \sigma_{ce} \tilde{\nabla}_b \sigma^{de} \left(\frac{1}{4}\sigma^{ab}\delta^c_d -\frac{1}{2} \sigma^{cb} \delta^a_d\right)+ \frac{D-2}{4}\sigma^{ab}\Omega_a \Omega_b\right]\label{eq:EHsimpcurved},
\ee
where $\tilde{\nabla}_a$ is the covariant derivative defined with respect to the background metric.   
Here we have defined $\Omega_a = \omega_a - \tilde{\omega}_a$, where $\tilde{\omega}_a = \partial_a \log \sqrt{-\tilde{g}}$.  In terms of the new variables, this quantity is 
\be
\Omega_a = \frac{1}{D-2}\sigma_{bc}\tilde{\nabla}_a \sigma^{bc} .
\ee 
Note that we have not added a matter action in \Eq{eq:EHsimpcurved}, so the background spacetime is Ricci-flat, i.e., $\tilde R_{ab} =0$.  However, extending our results to include matter is straightforward.  In particular, one simply adds $\sqrt{-\tilde{g}}\, \tilde{R}_{ab} \sigma^{ab}/16\pi G + \sqrt{-g}\, \mathcal{L}_\mathrm{matt}$ to the action and carries these terms through the equations of motion.  In any case, we will neglect matter hereafter.

Repeating our earlier analysis with the curved spacetime action in \Eq{eq:EHsimpcurved}, we generalize the cubic representations of the EH action in Eqs.~\eqref{eq:cubicA1} and \eqref{eq:cubicX}.  This is achieved by applying the replacement rules
\be 
\begin{aligned}
\eta_{ab} &\rightarrow \tilde{g}_{ab} \\
\partial_a &\rightarrow  \tilde{\nabla}_a\\
{\cal L}_{\rm EH} &\rightarrow  \sqrt{-\tilde g}\, {\cal L}_{\rm EH} . \label{eq:replacement}
\end{aligned}
\ee
In turn, the equation of motion for the auxiliary field $A^a_{bc}$ is the same as the flat space expression in \Eq{eq:eomChristoffel} except with the replacement $\Gamma^a_{bc} \rightarrow \Gamma^a_{bc} - \tilde{\Gamma}^a_{bc}$, where $\tilde{\Gamma}^a_{bc}$ is the background Christoffel symbol.  Similarly, the equation of motion for $B^a_{bc}$ sets this auxiliary field equal to $\Gamma^a_{bc} -\tilde{\Gamma}^a_{bc}$.

Meanwhile, the curved spacetime version of the simplified EH action in \Eq{eq:SEH_K} involves the generalization of the gauge-fixing term in \Eq{eq:SGF},
\be 
S_{\rm GF} = - \frac{1}{16\pi G}\int \mathrm{d}^D x \, \sqrt{-\tilde{g}}\, \frac{D-2}{4}\sigma^{ab} (\Omega_a - \mathcal{T}_a)(\Omega_b - \mathcal{T}_b),\label{eq:SGFcurved}
\ee
where we have defined the analogue of \Eq{eq:tau} in curved spacetime,
\be
\mathcal{T}_a = z \sigma_{ab} \tilde{\nabla}_c \sigma^{bc}.
\ee
After gauge fixing, we obtain the curved spacetime version of the simplified graviton Lagrangian, which is simply given by Eqs.~\eqref{eq:SEH_K} and \eqref{eq:Kgen} after the replacement in \Eq{eq:replacement}. 

\section{Conclusions}

\label{sec:conclusions}

In this paper we have reformulated the EH action in various forms that simplify the mechanics of graviton perturbation theory.  To derive these new representations, we have exploited the freedom of choosing a field basis and gauge fixing.  Our main results are: {\it i}) a purely cubic action for gravity in \Eq{eq:cubicA1} and {\it ii}) a simplified action in \Eq{eq:SEH_K} in which all interactions are trivially related to the graviton kinetic term.  

Having computed several reformulations of the EH action, it is useful to compare them among each other and to other work, notably the twofold Lorentz invariant action in \Ref{twofold} and the double copy relating gravity amplitudes to gauge theory amplitudes \cite{KLT,BCJ}, and ask which prescription provides the simplest method of calculation. One might be tempted to ask for a single formulation or action that is simplest for all computations in gravity, but in practice which version is most expedient to use depends on the nature of the calculation being done. For computing on-shell scattering of gravitons in a flat spacetime background, the double copy always wins, as explicit formulas for the gauge theory amplitudes are already known, so no new calculation is needed to compute graviton amplitudes. However, if one desires to find the off-shell currents, then the off-shell recursion relations derived in \Sec{sec:recursionrelations} are the best option. Note that such a simple recursion relation, in analogy with Berends-Giele recursion for Yang-Mills theory, was only possible because we were able to introduce auxiliary fields in \Sec{sec:cubic} to yield a cubic formulation of the EH action; in contrast, canonical perturbation theory or any of the other formulations of the EH action we derive without auxiliary fields contain new Feynman rules at each order in gravitons, significantly complicating any attempt to derive an off-shell recursion relation. 

Moreover, if one wishes to examine the gravitational equations of motion for perturbation theory to some fixed order about a curved spacetime background or using curvilinear coordinates, then it is possible that a formulation purely in terms of the perturbation $h$ may be most straightforward; such a calculation could be of use in astrophysical contexts for classical gravitational waves. In this case, the three candidates are the action derived in \Sec{sec:simp}, the twofold Lorentz invariant action derived in \Ref{twofold}, and the canonical perturbation expansion of the EH action. These actions can be compared by the number of terms they have at the first few orders in perturbation theory: at $\mathcal{O}(h^n)$ for $n =$ 2, 3, 4, 5, the canonical action has 4, 13, 35, 76 terms, the twofold Lorentz invariant action in \Ref{twofold} has 2, 2, 5, 5 terms in the Cayley-like basis and grows asymptotically as $3n^2/16$ terms, and the action in \Eqs{eq:SEH_K}{eq:Kgen} has exactly 5 terms at every order in perturbation theory, for all $n$. Thus, while the formulation of \Ref{twofold} is simpler at cubic order, the Lagrangian derived in the present paper eventually becomes simpler than any other known formulation of the EH action, allowing the Feynman vertex at arbitrary order in perturbation theory to be written explicitly in \Eq{eq:npointvertex}. However, there may be problems in classical gravity, outside of the context of perturbation theory, in which a first-order formulation of the equations of motion could be useful, as provided in \Eqs{eq:eomsigma}{eq:eomA} by our cubic action \eqref{eq:cubicA1}, which as noted previously is related to the Palatini formalism.

Our results leave several potential avenues for future work. First and foremost, it would be interesting to extend our results to higher loop order. As mentioned in text, the Jacobian associated with the field redefinition from $g^{ab}$ to $\sigma^{ab}$ is given in \Ref{twofold}. While we have restricted to tree-level scattering amplitudes in this paper, it should be straightforward to extend our results to loop level by introducing the appropriate Faddeev-Popov ghost.  It would be particularly interesting to construct a field basis in which these ghosts interact purely via cubic interactions. 

Second, because the Lagrangians in Eqs.~\eqref{eq:cubicA1} and \eqref{eq:SEH_K} are valid for arbitrary field excursions away from flat space, it should be possible to apply these representations to study classical curved spacetime backgrounds, e.g., the Schwarzschild solution. 

Last of all, the cubic structure and twofold Lorentz invariance of  \Eq{eq:cubicA1}  are strongly suggestive of the BCJ double copy.  It would be interesting to understand whether this is accidental or if there is indeed a direct connection between these ideas.

\begin{center} 
 {\bf Acknowledgments}
 \end{center}
 \noindent 
We thank Sean Carroll and Leo Stein for useful discussions and comments. C.C. is supported by a Sloan Research Fellowship and a DOE Early Career Award under Grant No. DE-SC0010255. G.N.R.~is supported by a Hertz Graduate Fellowship and a NSF Graduate Research Fellowship under Grant No.~DGE-1144469.

\bibliographystyle{utphys-modified}
\bibliography{GR_simplified}

\providecommand{\href}[2]{#2}\begingroup\raggedright\begin{thebibliography}{10}

\bibitem{LIGO}
{\bfseries LIGO Scientific Collaboration and Virgo} {\bfseries Collaboration},
  B.~P. Abbott { et~al.}, ``{Observation of Gravitational Waves from a Binary
  Black Hole Merger},''
  \href{http://dx.doi.org/10.1103/PhysRevLett.116.061102}{{\em Phys. Rev.
  Lett.} {\bfseries 116} (2016) 061102},
\href{http://arxiv.org/abs/1602.03837}{{\ttfamily arXiv:1602.03837 [gr-qc]}}.

\bibitem{Abbott:2016nmj}
{\bfseries LIGO Scientific Collaboration and Virgo} {\bfseries Collaboration},
  B.~P. Abbott { et~al.}, ``{GW151226: Observation of Gravitational Waves from
  a 22-Solar-Mass Binary Black Hole Coalescence},''
  \href{http://dx.doi.org/10.1103/PhysRevLett.116.241103}{{\em Phys. Rev.
  Lett.} {\bfseries 116} (2016) 241103},
\href{http://arxiv.org/abs/1606.04855}{{\ttfamily arXiv:1606.04855 [gr-qc]}}.

\bibitem{Abbott:2017vtc}
{\bfseries LIGO Scientific Collaboration and Virgo} {\bfseries Collaboration},
  B.~P. Abbott { et~al.}, ``{GW170104: Observation of a 50-Solar-Mass Binary
  Black Hole Coalescence at Redshift 0.2},''
  \href{http://dx.doi.org/10.1103/PhysRevLett.118.221101}{{\em Phys. Rev.
  Lett.} {\bfseries 118} (2017) 221101},
\href{http://arxiv.org/abs/1706.01812}{{\ttfamily arXiv:1706.01812 [gr-qc]}}.

\bibitem{Berends:1987me}
F.~A. Berends and W.~T. Giele, ``{Recursive Calculations for Processes with $n$
  Gluons},''
\href{http://dx.doi.org/10.1016/0550-3213(88)90442-7}{{\em Nucl. Phys.}
  {\bfseries B306} (1988) 759}.

\bibitem{Mafra:2015vca}
C.~R. Mafra and O.~Schlotterer, ``{Berends-Giele Recursions and the BCJ Duality
  in Superspace and Components},''
  \href{http://dx.doi.org/10.1007/JHEP03(2016)097}{{\em JHEP} {\bfseries 03}
  (2016) 097},
\href{http://arxiv.org/abs/1510.08846}{{\ttfamily arXiv:1510.08846 [hep-th]}}.

\bibitem{Ferraris1982}
M.~Ferraris, M.~Francaviglia, and C.~Reina, ``{Variational Formulation of
  General Relativity from 1915 to 1925 `Palatini's Method' Discovered by
  Einstein in 1925},'' \href{http://dx.doi.org/10.1007/BF00756060}{{\em General
  Relativity and Gravitation} {\bfseries 14} (1982) 243}.

\bibitem{Deser:1969wk}
S.~Deser, ``{Self-Interaction and Gauge Invariance},''
  \href{http://dx.doi.org/10.1007/BF00759198}{{\em Gen. Rel. Grav.} {\bfseries
  1} (1970) 9},
\href{http://arxiv.org/abs/gr-qc/0411023}{{\ttfamily arXiv:gr-qc/0411023
  [gr-qc]}}.

\bibitem{BjerrumBohr:2002kt}
N.~E.~J. Bjerrum-Bohr, J.~F. Donoghue, and B.~R. Holstein, ``{Quantum
  Gravitational Corrections to the Nonrelativistic Scattering Potential of Two
  Masses},'' \href{http://dx.doi.org/10.1103/PhysRevD.71.069903,
  10.1103/PhysRevD.67.084033}{{\em Phys. Rev.} {\bfseries D67} (2003) 084033},
  \href{http://arxiv.org/abs/hep-th/0211072}{{\ttfamily arXiv:hep-th/0211072
  [hep-th]}}.
[Erratum: Phys. Rev.D71,069903(2005)].

\bibitem{Neill:2013wsa}
D.~Neill and I.~Z. Rothstein, ``{Classical Space-Times from the S-Matrix},''
  \href{http://dx.doi.org/10.1016/j.nuclphysb.2013.09.007}{{\em Nucl. Phys.}
  {\bfseries B877} (2013) 177},
\href{http://arxiv.org/abs/1304.7263}{{\ttfamily arXiv:1304.7263 [hep-th]}}.

\bibitem{BCJ}
Z.~Bern, J.~J.~M. Carrasco, and H.~Johansson, ``{New Relations for Gauge-Theory
  Amplitudes},'' \href{http://dx.doi.org/10.1103/PhysRevD.78.085011}{{\em Phys.
  Rev.} {\bfseries D78} (2008) 085011},
\href{http://arxiv.org/abs/0805.3993}{{\ttfamily arXiv:0805.3993 [hep-ph]}}.

\bibitem{Monteiro:2014cda}
R.~Monteiro, D.~O'Connell, and C.~D. White, ``{Black Holes and the Double
  Copy},'' \href{http://dx.doi.org/10.1007/JHEP12(2014)056}{{\em JHEP}
  {\bfseries 12} (2014) 056},
\href{http://arxiv.org/abs/1410.0239}{{\ttfamily arXiv:1410.0239 [hep-th]}}.

\bibitem{Ridgway:2015fdl}
A.~K. Ridgway and M.~B. Wise, ``{Static Spherically Symmetric Kerr-Schild
  Metrics and Implications for the Classical Double Copy},''
  \href{http://dx.doi.org/10.1103/PhysRevD.94.044023}{{\em Phys. Rev.}
  {\bfseries D94} (2016) 044023},
\href{http://arxiv.org/abs/1512.02243}{{\ttfamily arXiv:1512.02243 [hep-th]}}.

\bibitem{Goldberger:2016iau}
W.~D. Goldberger and A.~K. Ridgway, ``{Radiation and the classical double copy
  for color charges},''
  \href{http://dx.doi.org/10.1103/PhysRevD.95.125010}{{\em Phys. Rev.}
  {\bfseries D95} (2017) 125010},
\href{http://arxiv.org/abs/1611.03493}{{\ttfamily arXiv:1611.03493 [hep-th]}}.

\bibitem{twofold}
C.~Cheung and G.~N. Remmen, ``{Twofold Symmetries of the Pure Gravity
  Action},'' \href{http://dx.doi.org/10.1007/JHEP01(2017)104}{{\em JHEP}
  {\bfseries 01} (2017) 104},
\href{http://arxiv.org/abs/1612.03927}{{\ttfamily arXiv:1612.03927 [hep-th]}}.

\bibitem{Parattu:2013gwa}
K.~Parattu, B.~R. Majhi, and T.~Padmanabhan, ``{Structure of the Gravitational
  Action and its Relation with Horizon Thermodynamics and Emergent Gravity
  Paradigm},'' \href{http://dx.doi.org/10.1103/PhysRevD.87.124011}{{\em Phys.
  Rev.} {\bfseries D87} (2013) 124011},
\href{http://arxiv.org/abs/1303.1535}{{\ttfamily arXiv:1303.1535 [gr-qc]}}.

\bibitem{Bern:2014sna}
Z.~Bern, S.~Davies, and T.~Dennen, ``{Enhanced Ultraviolet Cancellations in
  $\mathcal N=5$ Supergravity at Four Loops},''
  \href{http://dx.doi.org/10.1103/PhysRevD.90.105011}{{\em Phys. Rev.}
  {\bfseries D90} (2014) 105011},
\href{http://arxiv.org/abs/1409.3089}{{\ttfamily arXiv:1409.3089 [hep-th]}}.

\bibitem{Bern:2012gh}
Z.~Bern, S.~Davies, T.~Dennen, and Y.-t. Huang, ``{Ultraviolet Cancellations in
  Half-Maximal Supergravity as a Consequence of the Double-Copy Structure},''
  \href{http://dx.doi.org/10.1103/PhysRevD.86.105014}{{\em Phys. Rev.}
  {\bfseries D86} (2012) 105014},
\href{http://arxiv.org/abs/1209.2472}{{\ttfamily arXiv:1209.2472 [hep-th]}}.

\bibitem{Bern:2012cd}
Z.~Bern, S.~Davies, T.~Dennen, and Y.-t. Huang, ``{Absence of Three-Loop
  Four-Point Ultraviolet Divergences in $\mathcal{N}$=4 Supergravity},''
  \href{http://dx.doi.org/10.1103/PhysRevLett.108.201301}{{\em Phys. Rev.
  Lett.} {\bfseries 108} (2012) 201301},
\href{http://arxiv.org/abs/1202.3423}{{\ttfamily arXiv:1202.3423 [hep-th]}}.

\bibitem{Spradlin:2008bu}
M.~Spradlin, A.~Volovich, and C.~Wen, ``{Three Applications of a Bonus Relation
  for Gravity Amplitudes},''
  \href{http://dx.doi.org/10.1016/j.physletb.2009.02.059}{{\em Phys. Lett.}
  {\bfseries B674} (2009) 69},
\href{http://arxiv.org/abs/0812.4767}{{\ttfamily arXiv:0812.4767 [hep-th]}}.

\bibitem{ArkaniHamed:2008gz}
N.~Arkani-Hamed, F.~Cachazo, and J.~Kaplan, ``{What is the Simplest Quantum
  Field Theory?},'' \href{http://dx.doi.org/10.1007/JHEP09(2010)016}{{\em JHEP}
  {\bfseries 09} (2010) 016},
\href{http://arxiv.org/abs/0808.1446}{{\ttfamily arXiv:0808.1446 [hep-th]}}.

\bibitem{KLT}
H.~Kawai, D.~C. Lewellen, and S.~H.~H. Tye, ``{A Relation Between Tree
  Amplitudes of Closed and Open Strings},''
\href{http://dx.doi.org/10.1016/0550-3213(86)90362-7}{{\em Nucl. Phys.}
  {\bfseries B269} (1986) 1}.

\bibitem{Bern:1999ji}
Z.~Bern and A.~K. Grant, ``{Perturbative Gravity from QCD Amplitudes},''
  \href{http://dx.doi.org/10.1016/S0370-2693(99)00524-9}{{\em Phys. Lett.}
  {\bfseries B457} (1999) 23},
\href{http://arxiv.org/abs/hep-th/9904026}{{\ttfamily arXiv:hep-th/9904026
  [hep-th]}}.

\bibitem{Hohm:2011dz}
O.~Hohm, ``{On Factorizations in Perturbative Quantum Gravity},''
  \href{http://dx.doi.org/10.1007/JHEP04(2011)103}{{\em JHEP} {\bfseries 04}
  (2011) 103},
\href{http://arxiv.org/abs/1103.0032}{{\ttfamily arXiv:1103.0032 [hep-th]}}.

\bibitem{Cheung:2016prv}
C.~Cheung and C.-H. Shen, ``{Symmetry for Flavor-Kinematics Duality from an
  Action},'' \href{http://dx.doi.org/10.1103/PhysRevLett.118.121601}{{\em Phys.
  Rev. Lett.} {\bfseries 118} (2017) 121601},
\href{http://arxiv.org/abs/1612.00868}{{\ttfamily arXiv:1612.00868 [hep-th]}}.

\bibitem{Brandt}
F.~T. Brandt, J.~Frenkel, and D.~G.~C. McKeon, ``{General Covariant Gauge
  Fixing for Massless Spin-Two Fields},''
  \href{http://dx.doi.org/10.1103/PhysRevD.76.105029}{{\em Phys. Rev.}
  {\bfseries D76} (2007) 105029},
\href{http://arxiv.org/abs/0707.2590}{{\ttfamily arXiv:0707.2590 [hep-th]}}.

\end{thebibliography}\endgroup

\end{document}